\documentclass[artcl,twocolumn,showpacs,preprintnumbers,amsmath,amssymb,
floatfix,superscriptaddress]{revtex4-1}
\usepackage{graphics}
\usepackage{graphicx}
\usepackage{amsmath}
\usepackage{setspace}
\usepackage{braket}
\usepackage{epstopdf}
\usepackage{comment}
\usepackage{hyperref}
\usepackage{bm}
\usepackage{xfrac}
\usepackage{lineno}
\hypersetup{%
   pdfpagemode=None, %FullScreen,
   pdfstartpage=1,
   pdfmenubar=true,
   pdftoolbar=true,
   colorlinks = true,
   linkcolor=blue,
   citecolor=blue,
   urlcolor=blue,
   bookmarksopen=false
 }

\newcommand{\affA}{Van der Waals-Zeeman Institute, Institute of Physics,
University of Amsterdam, 1098 XH Amsterdam, Netherlands}
\newcommand{\affB}{Faculty of Physics, University of Warsaw, Pasteura 5,
02-093 Warsaw, Poland}

\begin{document}

\title{Buffer gas cooling of a trapped ion to the quantum regime}

\author{T.~Feldker}\affiliation{\affA}%\affiliation{\affE}
\author{H.~F\"urst}\affiliation{\affA}%\affiliation{\affE}
\author{H.~Hirzler}\affiliation{\affA}%\affiliation{\affE}
\author{N.~V.~Ewald}\affiliation{\affA}%\affiliation{\affE}
\author{M.~Mazzanti}\affiliation{\affA}%\affiliation{\affE}
\author{D.~Wiater}\affiliation{\affB}%\affiliation{\affW}
\author{M.~Tomza}\affiliation{\affB}%\affiliation{\affE}
\author{R.~Gerritsma$^*$}\affiliation{\affA}%\affiliation{\affE}\email{r.gerritsma@uva.nl}

\date{\today}

\begin{abstract}
Great advances in precision quantum measurement have been achieved with trapped ions and atomic gases at the lowest possible temperatures~\cite{Ludlow:2015,Bernien:2017,Zhang:2017}. These successes have
inspired ideas to merge the two systems~\cite{Tomza:2017cold}. In this way one can study the unique properties of ionic impurities inside a quantum fluid~\cite{Zipkes:2010,Schmid:2010,Ratschbacher:2013,Meir:2016, Kleinbach:2018,Sikorsky:2018, Haze:2018} or explore buffer gas cooling of the trapped ion quantum computer~\cite{Daley:2004}. Remarkably, in spite of its importance, experiments with atom-ion mixtures remained firmly confined to the classical collision regime~\cite{Schmid:2018}.
%It was suggested to reduce the collision energy by optimizing the ion-to-atom mass ratio~\cite{Cetina:2012}.
We report a collision energy of 1.15(0.23) times the $s$-wave energy (or 9.9(2.0)~$\mu$K) for a trapped ytterbium ion in an ultracold lithium gas. We observed a deviation from classical Langevin theory by studying the spin-exchange dynamics, indicating quantum behavior in the atom-ion collisions. Our results open up numerous opportunities, such as the exploration of atom-ion Feshbach resonances~\cite{Idziaszek:2011,Tomza:2015}, in analogy to neutral systems~\cite{Julienne:2010}.
\end{abstract}

\maketitle

%{\bf }

%reported collision energies of atom-ion mixtures have been at least two orders of magnitude higher than the $s$-wave energy~\cite{Schmid:2018}. %, while buffer gas cooling of trapped ions has been limited to even higher temperatures\,\cite{Haze:2018}.

Neutral buffer gas cooling of trapped ions has a long
history~\cite{Wesenberg:1995}, dating back to the times when laser cooling was still in its infancy.
The  development of atom trapping spurred efforts to employ quantum degenerate buffer gases. These are readily prepared in the 100~nK regime by evaporative cooling, making them superb coolants. Despite this, it is well-known that the time-dependent electric potential of a Paul trap complicates matters~\cite{Major:1968}. It causes a fast driven motion in the ions called
micromotion from which energy can be released when an ion
collides with an atom. This leads to a situation in which the kinetic energy of the ion becomes much larger than that of the surrounding buffer gas. This kept buffer gas cooling uncompetitive compared to laser cooling of the ions. It has also prevented the study of interacting ions and atoms in the quantum regime and reported collision energies of atom-ion mixtures have been at least two orders of magnitude higher than the $s$-wave energy~\cite{Schmid:2018}. It was suggested that this effect can be mitigated by employing an ion-atom combination with a large mass ratio~\cite{Cetina:2012} such as Yb$^+$ and $^6$Li. %, while buffer gas cooling of trapped ions has been limited to even higher temperatures\,\cite{Haze:2018}~\cite{Zipkes:2010, Schmid:2010, Haze:2018,Meir:2016, Tomza:2017cold, Kleinbach:2018}.

\begin{figure}
	%\centering
	\includegraphics[width=\columnwidth]{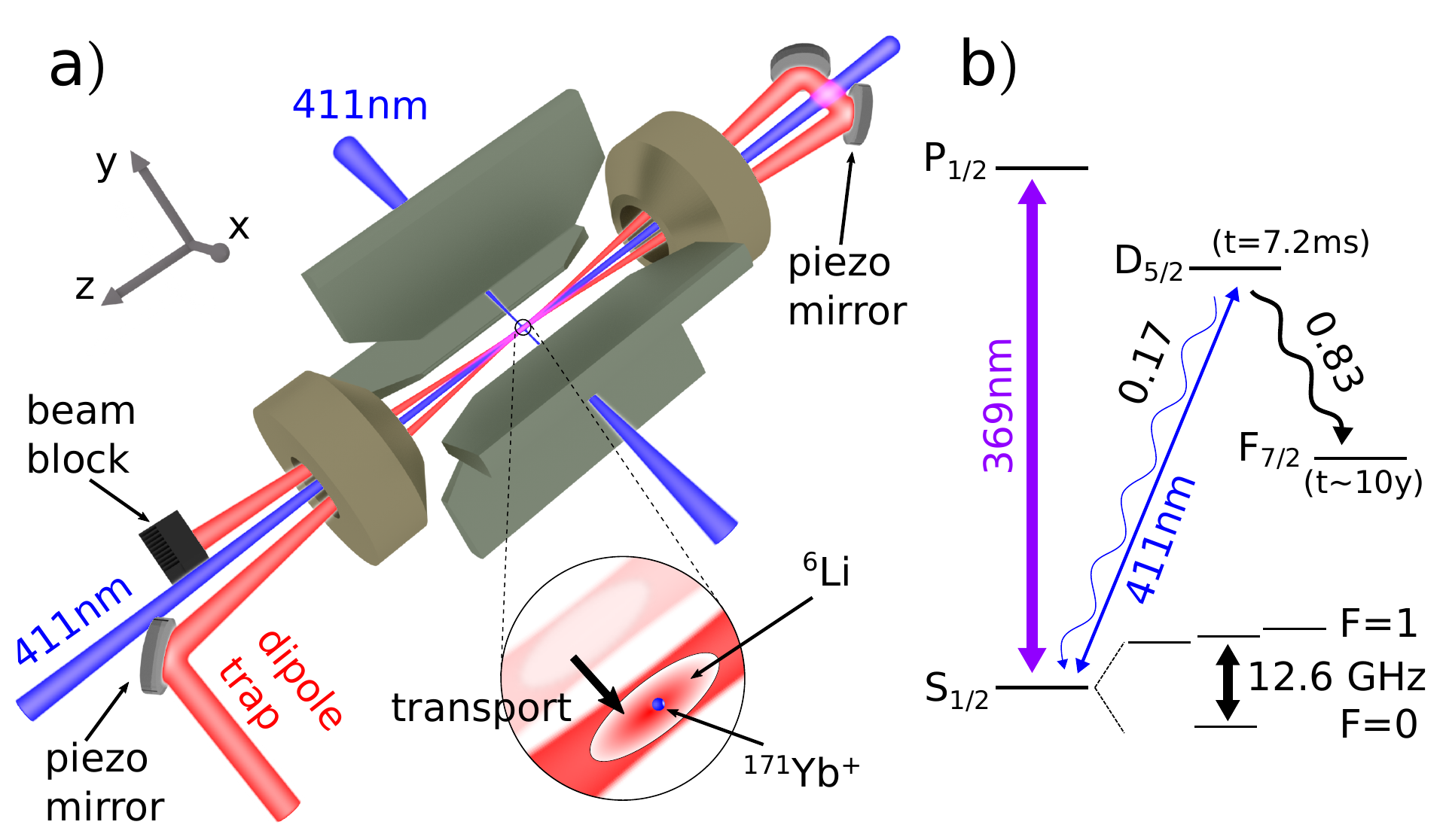}
  \caption{Setup. a) A cloud of ultracold $^6$Li atoms is prepared in an optical trap $\sim$~50~$\mu$m below a single ion in a Paul trap (shown in gray). The ion is
  then immersed in the atomic cloud by transporting the atom trap up using
  piezo-controlled mirrors. b) After a variable atom-ion interaction time, the
  ion is interrogated by coupling the $S_{1/2}$ ground state on a narrow
  quadrupole transition to the $D_{5/2}$ excited state. The coupling strength on the transition can be directly
  related to the temperature of the ion. State-selective fluorescence detection
  allows us to establish the average coupling strength.}
	\label{fig:setup}
\end{figure}

\begin{figure*}
	%\centering
	\includegraphics[width=1.5\columnwidth]{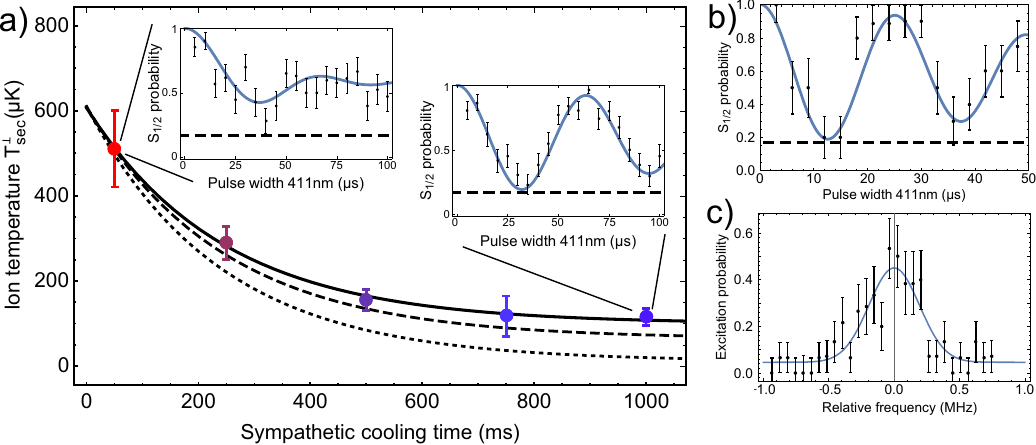}
	\caption{Cooling dynamics of an ion in the ultracold buffer gas. a) Ion temperature as a function of atom-ion interaction time together with an exponential fit (solid line) and molecular dynamics simulations with and without taking the time-dependence of the Paul trap into account (dashed and dotted lines, respectively). Before buffer gas cooling, the temperature of the ion is close to the Doppler laser cooling limit.
		This results in a rapidly decaying Rabi oscillation as a function of laser interrogation pulse width (inset). After buffer
		gas cooling, the Rabi oscillations persist for much longer. The branching
		ratio of decay out of the $D_{5/2}$ state limits the contrast of the Rabi oscillations to
		0.83 as indicated by the dashed horizontal lines. Measurement of radial ($T_{\text{sec}}^{\perp}=42(19)\,\mu$K)(b) and axial ($T_{\text{sec}}^{\text{ax}}=130(35)\,\mu$K)(c) temperatures
		after 1~s of interaction time with an atomic cloud with $T_{\text{a}} = $ 2.3(0.4)\,$\mu$K and after adiabatic decompression of the radial ion trap potential. Errorbars denote the quantum projection noise for the state detection, while errorbars for temperatures denote the standard deviation obtained from the fits.}
	\label{fig:cooling}
\end{figure*}

We trap and Doppler cool a single $^{171}$Yb$^+$ ion in
our Paul trap shown in Fig.~\ref{fig:setup}a) and prepare a cloud of
10$^4$~-~10$^5$ $^6$Li atoms with a temperature of $T_\text{a} = 2\,-\,10\,\mu$K in an
optical dipole trap 50~$\mu$m below the trapped ion.
The atoms are transported up by repositioning the dipole trap using
piezo-controlled mirrors. After a variable
interaction time, the ion is interrogated with a spectroscopy laser pulse at
411~nm that couples the $S_{1/2}$ ground state to the long-lived
$D_{5/2}$ state as shown in Fig.~\ref{fig:setup}b). We obtain the average kinetic energy of the ion by studying this laser excitation as a function of
pulse width~\cite{Leibfried:1996,Meir:2016}. In particular, the Rabi frequency
$\Omega$ of oscillations between the two states depends on the number of quanta $n$ present in the motion of the ion in its trap. Thermal occupation of excited states results in mixing of frequency
components and thus damping of the Rabi oscillation.
%Since the wavevector $\boldmath{k}$ of the interrogating beam has an angle of $45^\circ$ with respect to the radial modes of motion $x$ and $y$,
We fit the observed excitation to a model
that assumes a
%two-dimensional
thermal distribution with $\bar{n}$ motional quanta on average. %The trapfrequencies are $\omega_\text{x} \approx \omega_\text{y} = 330\,kHz$.
From this, we obtain the ion's secular temperature $T_{\text{sec}}^{\perp}\propto \bar{n}$ as explained in more detail in the Methods section.%, where we assume $T_{\text{sec}}^{\perp}=T_x=T_y$.

We observe buffer gas cooling of the ion by temperature measurements after
various hold times of the trapped ion in the ultracold cloud. The result for an
atomic cloud with temperature $T_\text{a} = $10\,$\mu$K and peak density
$\rho = 31(15) \times 10^{15}$\,m$^{-3}$ is shown in Fig.~\ref{fig:cooling}a).
Initially, the ion has a temperature of about
$T_{\text{sec}}^{\perp}=\,$600~$\mu$K, which is close to the Doppler cooling limit. Then, the ion
cools down with a $1/e$ time of $\tau_{\text{cool}}$ = 244(24)~ms to a final temperature of
$T_{\text{sec}}^{\perp}=$~98(11)~$\mu$K, corresponding to a mean number of motional quanta $\bar{n}=5.8(0.7)$ in the radial directions of motion. The buffer gas cooling thus outperforms Doppler cooling by about a factor of 5 in terms of attained temperature.

Note that the final ion temperature in Fig.~\ref{fig:cooling}a) is about an order of magnitude larger than the temperature of the buffer gas. This behaviour may be a direct consequence of the time-dependence of the ionic trapping potential~\cite{DeVoe:2009,Zipkes:2011,Chen:2014,Rouse:2017,Weckesser:2015,Meir:2016}, that causes energy release from the ion's micromotion during a collision. We investigate this by comparing the observed dynamics of the ion in the buffer gas with
classical molecular dynamics simulations~\cite{Fuerst:2018} in which we draw the ion's initial secular energy from a Maxwell-Boltzmann distribution at $T_\text{sec} = 609\,\mu$K to match the data.

If we run the simulations assuming a static ion trapping potential for the ion (this is known as the secular approximation~\cite{Berkeland:1998}), we find complete thermalization, $T_\text{sec}\rightarrow T_\text{a}$ as shown by the dotted line in Fig.~\ref{fig:cooling}a). We improve our model by including the time-dependence of the Paul trap using
parameters obtained from our experiment including all sources of micromotion as explained in the methods section.
In this simulation, a final ion temperature of 43\,$\mu$K is reached. When we also include the background heating rate of 85(50)\,$\mu$K/s that was measured in the absence of the atoms, the simulated final ion temperature reaches 63(12)\,$\mu$K, as shown by the dashed line in Fig.~\ref{fig:cooling}a).
%The simulations reproduce the buffer gas cooling time $\tau_{\text{cool}}$ if we set the simulated atomic density to $\rho_\text{at}=24(3)\times 10^{15}$\,m$^{-3}$, which is in agreement with the results from absorption imaging.
A likely explanation for the remaining discrepancy in final temperature is overestimation of the ion's kinetic energy by neglecting other dephasing mechanisms in the Rabi oscillations, such as laser frequency noise. Quantum corrections may also play a role at the small energies we obtain~\cite{Krych:2013}. We conclude that the temperature of the trapped ion is limited both by micromotion-induced heating during atom-ion collisions as well as the background heating rate of the ion.

%We include the long-range interaction potential between the atoms and the ion in the simulation since this is expected to play an important role at very low energies~\cite{Cetina:2012,Meir:2016}.

%In contrast to previous studies, we do not find a significant deviation from a thermal distribution after buffer gas cooling the ion in the simulations. We attribute this result to the large mass ratio of our ion-atom combination~\cite{DeVoe:2009,Zipkes:2011,Chen:2014,Rouse:2017,Weckesser:2015,Meir:2016}.
%More details can be found in the methods section.

To reach even lower energies in the experiment, we cool the atoms to $T_\text{a}$=2.3(0.4)\,$\mu$K and adiabatically lower the radial trap frequency for the ion from $\omega_x \approx \omega_y = 2\pi \times$ 330\,kHz to $\omega_x \approx \omega_y = 2\pi \times$ 210\,kHz at the end of 1\,s of buffer gas cooling. In this way we achieve a temperature of $T_{\text{sec}}^{\perp}=$~42(19)~$\mu$K corresponding to $\bar{n}=3.7(1.4)$ as depicted in Fig.~\ref{fig:cooling}b).

%Besides secular motion, the ion undergoes rapid micromotion due to the time-dependent electric fields in the Paul trap.
The total kinetic energy of the ion can be written as
$E_\text{i}=E_\text{sec}+E_\text{iMM}+E_\text{eMM}$, that is the secular (sec)
energy and the energy due to the intrinsic (iMM) and excess micromotion (eMM).
%Since the light propagates at 45$^{\circ}$ with respect to $x$ and $y$, we obtain the average secular energy in both radial directions.
To obtain the total collision energy we have to additionally determine the axial secular temperature and the various micromotion energies.

Due to the weak confinement along the trap axis ($\omega_z =$ 2$\pi \times$ 130\,kHz), it is more
convenient to probe the excitation probability as a function of the frequency of
the laser, which we now direct along the $z$-axis. Thermal motion leads to Doppler broadening of the resonance as plotted in Fig.~\ref{fig:cooling}c). %We use a short pulse width of 5\,$\mu$s in order to not resolve the motional sidebands.
We fit a Gaussian distribution to the data and we find $\sigma = 193(26)$\,kHz, corresponding to
$T_{\text{sec}}^{\text{ax}}$~=~130(35)~$\mu$K. The larger value compared to $T_{\text{sec}}^{\perp}$ has two reasons: Firstly, the weaker axial trap potential gives rise to a higher background heating rate (200\,$\mu$K/s) and thus limits the attainable final temperature, and secondly the thermometry method is less reliable and more prone to overestimation of the temperature due to saturation broadening.

Intrinsic micromotion leads to a kinetic
energy $E_\text{iMM}\approx k_{\text{B}}T_{\text{sec}}^{\perp}$~\cite{Berkeland:1998}.
Excess micromotion occurs because of experimental imperfections that modify the trap potential.
Details on the compensation and characterization of excess micromotion can be
found in the methods section. In the experiment, we
find $E_\text{eMM}/k_\text{B}\leq $~44(13)~$\mu$K.

The collision energy is given by~\cite{Fuerst:2018}:
\begin{equation}
E_\text{col}=\frac{\mu}{m_\text{i}}E_\text{i}+\frac{\mu}{m_\text{a}}E_\text{a},
\end{equation}
with $m_\text{i}$ and $m_\text{a}$ the mass of the ion and atom, respectively,
$\mu$ the reduced mass and $E_\text{a}=3k_\text{B}T_\text{a}/2$ the average kinetic
energy of the atoms. Note that due to the large mass
ratio, $\mu\approx m_\text{a}\ll m_\text{i}$. Taking into account the contribution of all types of motion as summarized in Table\,1 results in a collision energy of $E_\text{col}=1.15(0.23)\times
E_s$, with $E_s/k_\text{B} =8.6$~$\mu$K the $s$-wave collision
energy~\cite{Tomza:2017cold}.

\begin{table}[htbp]
	\centering
	\begin{tabular}{ccc}\\
		\hline
		
		\multicolumn{1}{|l|}{\textbf{Type of motion}} & \multicolumn{1}{l|}{$E_{\text{kin}}/\text{k}_\text{B}(\mu \text{K})$} & \multicolumn{1}{l|}{$E_{\text{col}}/\text{k}_\text{B}(\mu \text{K})$} \\
	
		\hline
		\multicolumn{1}{|l|}{Radial secular ion} &\multicolumn{1}{c|}{$2 \times 21(9)$} & \multicolumn{1}{c|}{$1.4(0.6)$} \\  \hline
		\multicolumn{1}{|l|}{Intrinsic micromotion} &\multicolumn{1}{c|}{$2 \times 21(9)$} & \multicolumn{1}{c|}{$1.4(0.6)$}  \\  \hline	
		\multicolumn{1}{|l|}{Axial secular ion} &\multicolumn{1}{c|}{$65(18)$} & \multicolumn{1}{c|}{$2.2(0.4)$}  \\  \hline
		\multicolumn{1}{|l|}{Excess micromotion} &\multicolumn{1}{c|}{$44(13)$} & \multicolumn{1}{c|}{$1.5(0.4)$}  \\  \hline	
		\multicolumn{1}{|l|}{\textbf{Total ion energy}} &\multicolumn{1}{c|}{\textbf{193(42)}} & \multicolumn{1}{c|}{\textbf{6.6(1.4)}}  \\  \hline	
		\multicolumn{1}{|l|}{Atom temperature} &\multicolumn{1}{c|}{$3/2 \times 2.3(0.4)$} & \multicolumn{1}{c|}{$3.3(0.6)$}  \\  \hline
		\multicolumn{1}{|l|}{\textbf{Total collision energy}} &\multicolumn{1}{c|}{---} & \multicolumn{1}{c|}{\textbf{9.9(2.0)}}  \\  \hline	
	\end{tabular}
	\caption{Measured energy budget of the trapped ion and atoms in terms of kinetic energy ($E_{\text{kin}}$) and collision energy ($E_{\text{col}}$). Errors are given in units of $\mu$K.}

\end{table}

\begin{figure}[t]
	%\centering
	\includegraphics[width=\columnwidth]{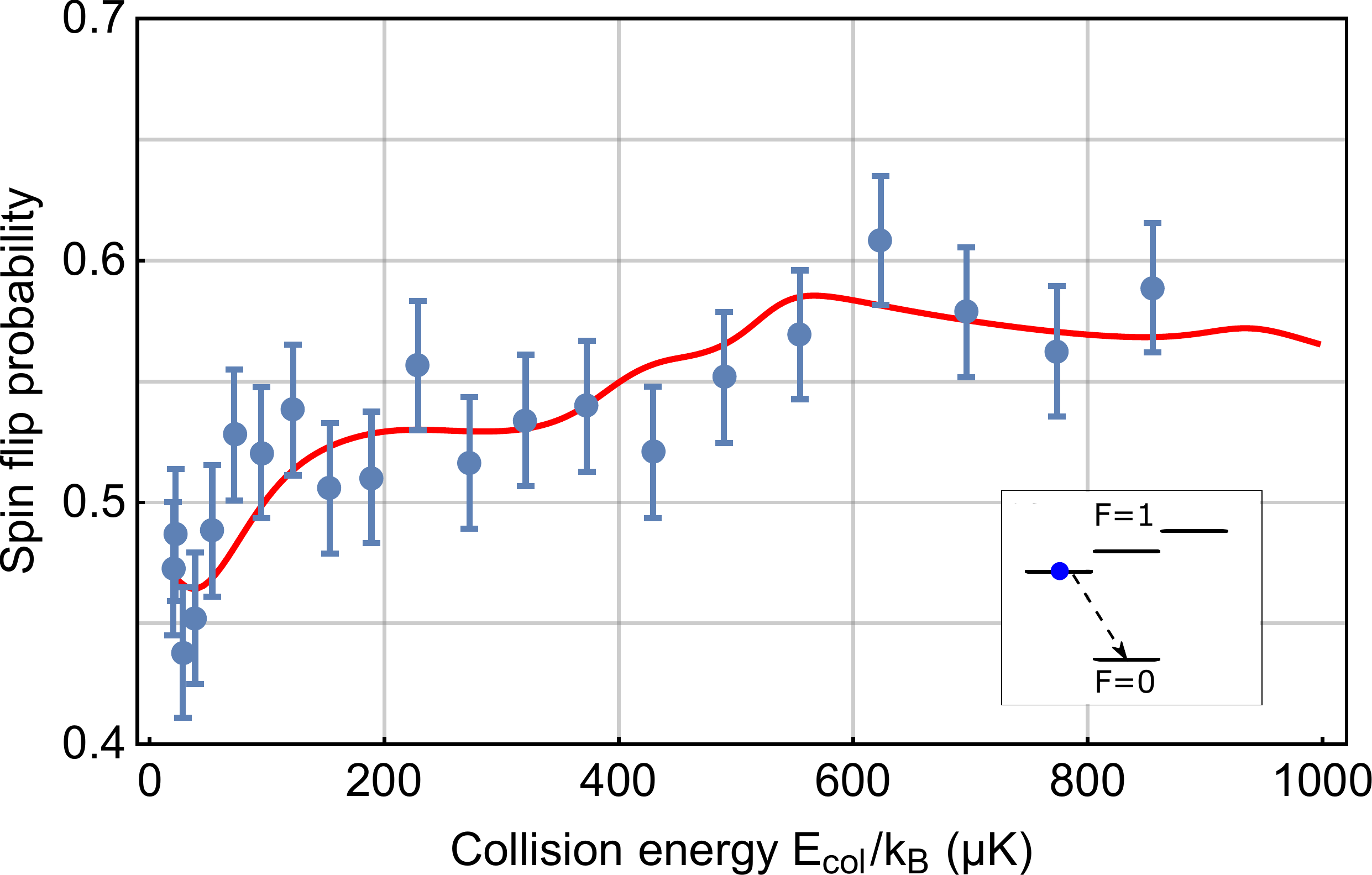}
  \caption{Spin-exchange rate versus collsion energy. Probability of detecting the ion spin in $|F=0,m_F=0\rangle$ after
  preparing it in $|F=1,m_F=-1\rangle$ and letting it interact with the atomic
  cloud for about 10~ms. The collision energy is varied
  using an offset field to tune the excess micromotion energy $E_{\text{eMM}}$.
  %On the horizontal axis we denote the collision energy corresponding to the maximum of the coherent energy distribution (plotted in inset a)) added to a thermal offset of 20\,$\mu$K.
  %The dashed line is based on quantum scattering calculations for sharp collision energies.
  The red solid line is obtained by convolving the rate from quantum scattering calculations with the energy distribution of the ion as explained in the methods section. Both theory and data show
  a clear deviation from classical behavior in which the spin-exchange rate should be independent of collision energy. The inset shows the measured spin exchange process in the ion. Errobars denote the quantum projection noise.}
	\label{fig:spin}
\end{figure}
Now that we cooled the mixture close to the $s$-wave limit we expect the atom-ion interactions to be governed by quantum mechanics. To look for signs of quantum behavior in the interaction we investigate the occurrence of spin-changing collisions~\cite{Ratschbacher:2013,Sikorsky:2018,Fuerst:2018:spin} as a function of collision energy. Spin-exchange is associated with short-range collisions between the atoms and
ions, known as Langevin collisions. In the classical regime, the Langevin
collision rate is strictly independent of collision energy~\cite{Tomza:2017cold}. At very low
collision energy, however, the quantization of the collision angular momentum and quantum reflection
start to play a role. This leads to the occurrence of structure such as shape resonances in the spin-exchange
rate. The details of this structure depend on the singlet and triplet
scattering lengths that quantify the interactions between the atom and ion in the quantum regime.

%The attained low collision energies should result in quantum effects in the
%interactions. To find these, we investigate the occurrence
%of spin-changing collisions as a function of collision energy~\cite{Ratschbacher:2013,Sikorsky:2018,Fuerst:2018:spin}.
%Spin-exchange is associated with short-range collisions between the atoms and
%ions, known as Langevin collisions. In the classical regime, the Langevin
%collision rate is strictly independent of collision energy~\cite{Tomza:2017cold}. At very low
%collision energy, however, the quantization of the collision angular momentum and quantum reflection
%start to play a role. This leads to the occurrence of structure such as shape resonances in the spin-exchange
%rate. The details of this structure depend on the singlet and triplet
%scattering lengths that quantify the interactions between the atom and ion in the quantum regime.

After buffer gas cooling for 1~s in an atomic cloud with $T_{\text{a}} = 10\,\mu$K, we prepare the ion in the state $|S_{1/2},F=1,m_F=-1\rangle$, with a microwave pulse. Here $F$ denotes the total angular momentum and $m_F$ its projection on the quantization axis. The atomic ensemble is in
a spin-mixture of the lowest two Zeeman states $|S_{1/2},F=1/2,m_F=\pm1/2\rangle$. Due to spin-exchange collisions during the interaction time the ion can fall to the $|S_{1/2},F=0,m_F=0\rangle$ state. We let the ion
interact with the cloud of atoms with a density of 21(10)$\times 10^{15}$\,m$^{-3}$ for about 10~ms, corresponding to about one Langevin
collision.
Only during the interaction time, we give the ion
a variable amount of excess micromotion energy of $E_{\text{eMM}}$ by ramping
offset voltages on compensation electrodes~\cite{Tomza:2017cold,Joger:2017}. We then shelve the population that remains in the $|S_{1/2},F=1\rangle$ state to the long lived $F_{7/2}$ state. Subsequent fluorescence detection allows to discriminate between an ion in the $|S_{1/2},F=0\rangle$ state (spin exchange) and an ion in the $F_{7/2}$ state (no spin exchange) with near unit fidelity. In
Fig.~\ref{fig:spin} the result of averaging 309~of such experimental runs is
shown. We see a significant dependence of the spin-exchange rate on the collision energy and thus a clear deviation from the classical prediction especially for low collision energies.

To gain further insight into the data, we compare it to multi-channel quantum scattering calculations based on the complete description of molecular and hyperfine structures. The amplitude, slope, and shape of the rate constants in the investigated energy range depend strongly on the values of the singlet and triplet
scattering lengths. In Fig.~\ref{fig:spin} the calculated rate constants for the spin-exchange collisions convoluted with the experimental collision energy distribution are presented for the singlet and triplet scattering lengths of $a_\text{S} = 1.2(0.3)R_4$ and $a_\text{T} = -1.5(0.7)R_4$ with $R_4 = 70$~nm~\cite{Tomza:2017cold} and 1.2(0.4) Langevin collisions on average during the interaction time. These values provide the best fit to the experimental data. %The large difference between the obtained scattering lengths agrees with our previous finding~\cite{Joger:2017} based on the phase locking mechanism in the classical regime~\cite{Sikorsky:2018lock,Cote:2018}.

In conclusion, we have demonstrated buffer gas cooling of a single ion in a Paul
trap to the quantum regime of atom-ion collisions.
This has been an elusive goal in hybrid atom-ion
experiments for more than a decade~\cite{Tomza:2017cold}. The data and simulations suggest that even lower temperatures may be reached when using colder and denser atomic clouds, both of which are technically feasible. In particular, a denser cloud would allow eliminating the background heating rate of the ion. We speculate that controlling elastic atom-ion collisions using possible Feshbach resonances~\cite{Idziaszek:2011,Tomza:2015} may allow tuning the cooling rate and accessible temperatures in atom-ion mixtures further, as is the case in neutral systems~\cite{Julienne:2010}.

%We measured the spin exchange rate versus the collision energy and found significant deviations from the
%classical prediction of a temperature independent rate as we approach the $s$-wave limit. We obtain a first estimate for the singlet and triplet scattering lengths of the $^6$Li-$^{171}$Yb$^+$ combination from a comparison of our data to quantum scattering calculations.
%Our results open up numerous opportunities, such as the possible
%observation of Feshbach resonances between atoms and ions~\cite{Idziaszek:2011,Tomza:2015}, the study of
%atom-ion chemistry in the quantum regime and the usage of the ultracold buffer
%gas as a coolant for the trapped ion quantum computer~\cite{Daley:2004}.

%\appendix

\section{Methods}

\subsection{Determination of ion energy}

To obtain information on its motional state, the ion is interrogated with a spectroscopy laser pulse at
411~nm that couples the $|S_{1/2},F=0,m_F=0\rangle$ state to the
$|D_{5/2},F'=2,m_F'=0\rangle$ state (see Fig.~\ref{fig:setup}). This state will decay to the long-lived
$F_{7/2}$ state or back to the ground state with probabilities of 0.83 and 0.17,
respectively~\cite{Taylor:1997}. Subsequent fluorescence imaging allows us to detect these states with near unit fidelity. The Rabi frequency
$\Omega$ of oscillations on the spectroscopy transition depends on the amount of
motional quanta $n_i$ present in the secular motion of the ion according to
$\Omega= \Omega_0\prod_{i=x,y,z} e^{-\eta_i/2}L_{n_i}(\eta_i^2)$~\cite{Leibfried:1996,Meir:2016}, with the ground state Rabi frequency
$\Omega_0$, Lamb-Dicke parameter $\eta_i=k_il^{\rm ho}$, $k_i$ the wavevector of the
411~nm light projected onto the direction of ion motion $i$ and $l^{\rm
ho}=\sqrt{\hbar/(2m_{\rm i}\omega_i)}$ the size of the ionic groundstate
wavepacket. Here, $L_{n_i}(\eta_i^2)$ denotes the Laguerre polynomial. Since the laser beam has a 45$^{\circ}$ angle with respect to the $x$- and $y$ direction of ion motion and $\omega_x\approx \omega_y=\omega_{\perp}$, we set $\eta_x=\eta_y=\eta/\sqrt{2}$ and $\eta_z=0$ for the measurements on the radial motion. Thermal occupation of excited states results in mixing of frequency
components and thus damping of the Rabi oscillation. To obtain the ion
temperature $T_{\text{sec}}^{\perp}=\hbar\omega(\bar{n}+1/2)/k_\text{B}$ with $\bar{n}=(\bar{n}_x+\bar{n}_y)/2$ the average number of motional quanta, we fit the observed excitation to a model
that assumes a thermal distribution with $P_{\bar{n}_{x,y}}(n)=\bar{n}_{x,y}^n/(1+\bar{n}_{x,y})^{n+1}$ for each direction of motion $x$ and $y$. Here, we assume $\bar{n}_x$=$\bar{n}_y$.

\subsection{Micromotion compensation}

\begin{figure}
	%\centering
	\includegraphics[width=\columnwidth]{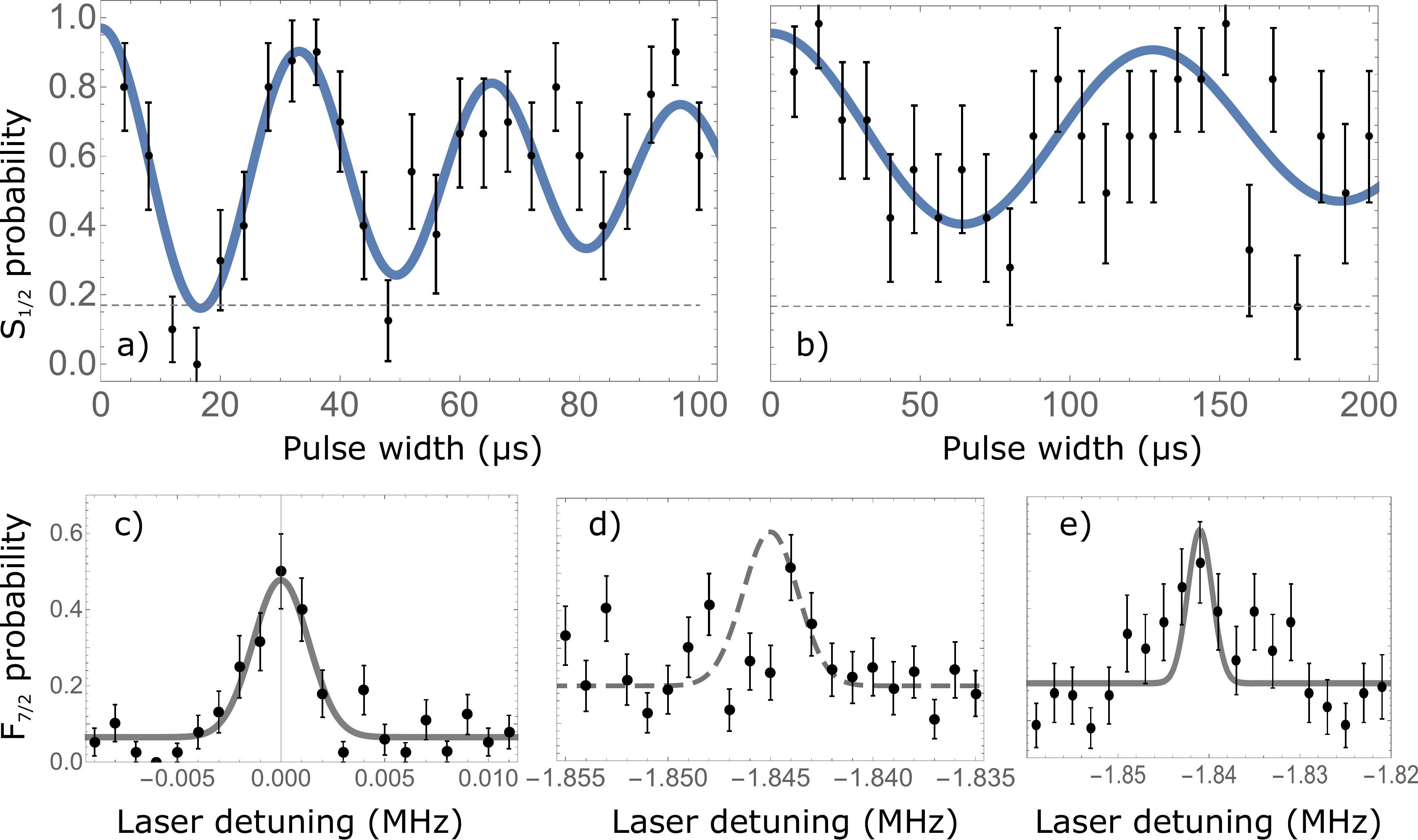}
  \caption{Micromotion analysis with resolved sideband spectroscopy. In part a) and b) Rabi oscillations on the carrier and the
  micromotion sideband for optimal compensation settings are plotted. From
  a comparison of the Rabi frequencies $\Omega_{\text{car}}=2\pi \times
  32.0(0.8)\,$kHz and $\Omega_{\text{MM}}=2\pi \times 7.0(0.5)\,$kHz in
  combination with the applied laser powers of $P_{411} = 32\,\mu$W and $P_{411}
  = 840\,\mu$K, respectively we obtain a residual micromotion energy of
  $\bar{E}_{\text{eMM}}/k_{\text{B}} = 21.5(1.5)\,\mu$K. Part c) shows
  a frequency scan over the carrier transition, carried out with a laser power
  of $P_{411} = 61\,\mu$W. A clear peak is visible. For the data plotted in part
  d) the frequency of the laser is shifted by $-\Omega_{\text{rf}} = -1.85\,$MHz
  compared to c) and the power is increased to $P_{411} = 21.7\,$mW. At the
  expected resonance frequency for the micromotion sideband we do not see
  a clear peak, only the background is higher compared to c) due to off-resonant
  carrier excitation at these high laser powers. If we shift the ion out of the
  optimal position for minimal micromotion we observe a clear resonance again as
  plotted in e). We conclude that the Rabi frequency $\Omega_{\text{MM}}$
  on the micromotion sideband presented in e) is not larger than the Rabi
  frequency on the carrier $\Omega_{\text{car}}$ presented in c). From this we
  obtain an upper limit of the axial micromotion at the optimal position of
  $\bar{E}_{\text{eMM}}/k_{\text{B}} = 33\,\mu$K.}
	\label{fig:MM}
\end{figure}

The Paul trap operates at a drive frequency of
$\Omega_{\text{rf}}=2\pi\times$~1.85~MHz. The motion of an ion in the trap is composed of a secular part with eigenfrequencies $\omega_x,\omega_y,\omega_z$ and
its intrinsic and excess micromotion at the drive frequency $\Omega_{\text{rf}}$. Intrinsic micromotion cannot be avoided and leads to an additional kinetic energy on the order of $E_\text{iMM}\sim k_BT_{\text{sec}}^{\perp}$~\cite{Berkeland:1998}. Buffer gas cooling to ultracold temperatures requires excellent control over
excess micromotion. Not only does excess micromotion hinder cooling to the
lowest secular temperatures as energy from the fast driven micromotion can be
transferred to the secular motion of the ion during a collision with an ultracold
atom, but the kinetic energy stored in the micromotion increases the overall
collision energy. In the following we describe our methods to compensate
micromotion to the required level. Furthermore we give a detailed analysis of
the micromotion energy budget.

\paragraph{Stray fields}
The primary cause for excess micromotion are stray electric fields, shifting the
ion out of the rf-quadrupole node. We determine the remaining stray electric
fields and the resulting excess micromotion with a set of two complementary
methods. In horizontal direction we obtain the dc-electric field  by measuring
the ions position by florescence imaging on a camera as a function of radial
trapping potential $\omega_{\text{rad}}$. The position shift of the ion in an
electric field $E_{\text{DC}}$ is given by

\begin{equation}\label{x_vs_Edc}
x(\omega_{\text{rad}}) = E_{\text{DC}}\times \frac{e}{m_{\text{Yb}}}\times \omega_{\text{rad}}^{-2}
\end{equation}

\noindent with $e$ denoting the elementary charge and $m_{\text{Yb}}$ the mass
of the Yb$^+$ ion. Fitting the data, we obtain a stray field of $E_{\text{DC}}
= 10(10)\,$mV/m. In order to account for drifts between micromotion compensation
measurements we assume a slightly higher limit of $E_{\text{DC}} \leq 50\,$mV/m.
The average micromotion energy $\bar{E}_{\text{eMM}}$ is calculated as

\begin{equation}\label{EeMM}
\bar{E}_{\text{eMM}}(E_{\text{DC}}) = \frac{E_{\text{DC}}^2 \times e^2}{2 m_{\text{Yb}} \times \omega_{\text{rad}}^2},
\end{equation}

\noindent resulting in an excess micromotion energy of
$\bar{E}_{\text{eMM}}/k_{\text{B}} \leq 4.7\,\mu$K
($\bar{E}_{\text{eMM}}/k_{\text{B}} \leq 1.9\,\mu$K) for a radial potential of
$\omega_{\text{rad}} = 2\pi \times 210$\,kHz ($\omega_{\text{rad}} = 2\pi \times
330$\,kHz).

In vertical direction we measure stray fields using microwave Ramsey
spectroscopy on the ($^2S_{1/2}, F=0) \leftrightarrow (^2S_{1/2}, F=1,
m_F=1$) transition. We apply a magnetic field gradient of $g_v = 0.17$\,T/m
leading to a frequency shift of the transition by 2.3\,kHz/$\mu$m. We determine
the ion shift for radial potentials of $\omega_{\text{rad}} = 2\pi \times
25\,$kHz and $\omega_{\text{rad}} = 2\pi \times 330\,$kHz. From a linear fit to
the measured frequency shifts we obtain the required compensation voltage
$V_{\text{comp}}$ with an uncertainty of 0.05\,V. To account for daily drifts we
assume a miscompensation of $V_{\text{comp}} \leq$ 0.2\,V. We obtain the
micromotion energy due to this miscompensation by calibrating the energy scale
with resolved sideband spectroscopy on the narrow $^2S_{1/2} \leftrightarrow
^2D_{5/2}$ transition as explained below. For a radial potential of
$\omega_{\text{rad}} = 2\pi \times 210$\,kHz, we obtain
$\bar{E}_{\text{eMM}}/k_{\text{B}} \leq 8.3\,\mu$K.

\paragraph{Energy calibration}

\begin{table*}[t!]
	\centering
	\begin{tabular}{ccc}\\
		\hline
		
		\multicolumn{1}{|l|}{\textbf{Type of micromotion}} & \multicolumn{1}{l|}{$\bar{E}_{\text{eMM}}(\text{210kHz})/k_{\text{B}}(\mu \text{K})$} & \multicolumn{1}{l|}{$\bar{E}_{\text{eMM}}(\text{330kHz})/k_{\text{B}}(\mu \text{K})$} \\
		
		\hline
		\multicolumn{1}{|l|}{Axial} &\multicolumn{1}{c|}{$13(13)$} & \multicolumn{1}{c|}{$33(33) $} \\  \hline
		\multicolumn{1}{|l|}{Radial Quadrature} &\multicolumn{1}{c|}{$2 \times 8.7 (0.6)$} & \multicolumn{1}{c|}{$2 \times 21.5(1.5) $}  \\  \hline	
		\multicolumn{1}{|l|}{Radial field (vertical)} &\multicolumn{1}{c|}{$ \leq 8.3$} & \multicolumn{1}{c|}{$ \leq 3.4 $}  \\  \hline
		\multicolumn{1}{|l|}{Radial field (horizonatal)} &\multicolumn{1}{c|}{$ \leq 4.7$} & \multicolumn{1}{c|}{$ \leq 2.1 $}  \\  \hline
		\multicolumn{1}{|l|}{Total} &\multicolumn{1}{c|}{$44(13)$} & \multicolumn{1}{c|}{$82(33) $}  \\  \hline	
		
	\end{tabular}
  \caption{Measured excess micromotion budget of the trapped ion at
  $\omega_{\text{rad}} = 210\,$kHz and $\omega_{\text{rad}} =330\,$kHz. The
  total obtained excess micromotion energy at $\omega_{\text{rad}} = 210$\,kHz
  is $\bar{E}_{\text{eMM}}/k_{\text{B}} = 44(13)\,\mu$K. The values in brackets
  denote the error in $\mu$K. The error is dominated by the error of the axial
  micromotion measurement.}\label{tab:MMlimits}
\end{table*}

We calibrate the excess micromotion energy versus Voltage on the compensation
electrodes by using resolved sideband spectroscopy on the $^2S_{1/2}
\leftrightarrow ^2D_{5/2}$ transition. We use the magnetic field insensitive
$(F=0,m_F=0) \leftrightarrow (F=2,m_F=0)$ transition in $^{171}$Yb$^+$. We compare
the Rabi frequency on the micromotion sideband and the carrier at
V$_{\text{comp}} =$7\,V. We obtain $\Omega_{\text{MM}}=2\pi \times
28.3(0.9)\,$kHz and $\Omega_{\text{car}}=2\pi \times 39.0(1.2)\,$kHz. Solving

\begin{equation}\label{key}
\frac{J_{0}(\beta_{\text{MM}})}{J_{1}(\beta_{\text{MM}})}=\frac{\Omega_{\text{car}}}{\Omega_{\text{MM}}}
\end{equation}

\noindent with $J_i$ denoting Bessel functions, yields a modulation index
$\beta_{\text{MM}} = 1.18$. From the modulation index we obtain the average
kinetic energy as

\begin{equation}\label{key}
\bar{E}_{\text{eMM}} = \frac{m}{4}\left(\frac{\beta_{\text{MM}}\times\Omega_{\text{rf}}}{k}\right)^2
\end{equation}

\noindent with $k$ the projection of the wavevector on the direction of the
micromotion and $\Omega_{\text{rf}} = 1.85\,$MHz the drive frequency of the
trap. This results in $\bar{E}_{\text{eMM}}/k_{\text{B}} = 84(7)\,\mu$K/V$^2$
$\times V_{\text{comp}}^2$ and $\bar{E}_{\text{eMM}}/k_{\text{B}}
= 208(19)\,\mu$K/V$^2$ $\times V_{\text{comp}}^2$ for radial potentials of
$\omega_{\text{rad}} = 2\pi \times 330$\,kHz and $\omega_{\text{rad}} = 2\pi
\times 210$\,kHz respectively.

\paragraph{Quadrature micromotion}

After carefully compensating any stray electric fields, we measure the remaining
micromotion by resolved sideband spectroscopy. We set the radial potential to
$\omega_{\text{rad}} = 2\pi \times 330$\,kHz. We compare the Rabi frequency on
the carrier, $\Omega_{\text{car}}=2\pi \times 32.0(0.8)\,$kHz at a laser power
of $P_{411} = 32\,\mu$W with the Rabi frequency on the micromotion sideband
$\Omega_{\text{MM}}=2\pi \times 7.0(0.5)\,$kHz at $P_{411} = 840\,\mu$W. We
obtain a micromotion energy of $\bar{E}_{\text{eMM}}/k_{\text{B}}
= 21.5(1.5)\,\mu$K. This value includes radial micromotion caused by remaining
stray electric fields as well as quadrature micromotion caused by a phase
difference of the rf-signal on the opposing rf-electrodes. Since we can not
differentiate between these types of radial micromotion, the obtained value is
an upper limit for quadrature micromotion. The laser beam propagates at an angle
of $\pi/4$ with respect to the direction of quadrature micromotion so that we
have to multiply the measured value by two in order to account for the full
micromotion energy. Quadrature micromotion energy is proportional to the square
of the trapping potential so that we get $\bar{E}_{\text{eMM}}/k_{\text{B}}
= 2\times8.7(0.6)\,\mu$K for $\omega_{\text{rad}} = 2\pi \times 210$\,kHz.

\paragraph{Axial micromotion}

Finite size effects of the linear Paul trap lead to a weak rf-potential in the
direction of the trap axis. Thus, the oscillating electric field vanishes at one
point on the axis only. We position the single ion in our trap to this point and
measure an upper limit to the remaining axial micromotion.  Axial micromotion
can in principle be measured in the same way as described for the quadrature
micromotion, using a laser beam propagating along the trap axis. However, since
our axial potential is rather weak  $\omega_{\text{ax}} \leq 2\pi \times
130$\,kHz, and the corresponding Lamb-Dicke parameter
$\eta_{\text{ax}}\geq0.23$, we do not observe coherent oscillation when exciting
with a laser beam propagating along the trap axis. In order to still measure an
upper limit for axial micromotion we compare a frequency scan over the carrier
at very low power $P_{411} = 61\,\mu$W with a scan over the micromotion sideband
at full power $P_{411} = 21.7\,$mW. From the data presented in Fig. 1 we see
that the transition on the micromotion sideband at these settings is not stronger than the
carrier transition. From the ratio of laser powers, we calculate an upper bound
to the axial micromotion of $\bar{E}_{\text{eMM}}/k_{\text{B}} = 33\,\mu$K for
$\omega_{\text{rad}} = 2\pi \times 330$\,kHz. If we reduce the radial trap
potential to $\omega_{\text{rad}} = 2\pi \times 210$\,kHz we obtain
$\bar{E}_{\text{eMM}}/k_{\text{B}} = 13\,\mu$K.  The obtained limits for
micromotion energy at $\omega_{\text{rad}} = 2\pi \times 210$\,kHz and
$\omega_{\text{rad}} = 2\pi \times 330$\,kHz are summarized in Table~\ref{tab:MMlimits}.

\subsection{Tuning the collision energy}

In the experiment, we tune the kinetic energy of the ion by shifting it out of
the Paul trap center with an electric control field $E_{\text{DC}}$. The
resulting micromotion experienced by the ion causes a coherent motion with an
energy distribution for $E\leq 2\bar{E}_{\text{eMM}}(E_{\text{DC}})$:

\begin{equation}
P_{\bar{E}_{\text{eMM}}}(E)=\frac{1}{\pi}\frac{1}{\sqrt{E(2\bar{E}_{\text{eMM}}(E_{\text{DC}})-E)}}.
\end{equation}

\noindent To compare the data to the quantum scattering calculations, we
convolute the calculated spin-exchange rates $\gamma(E)$ with this energy
distribution. Here, we assume a thermal offset of 20~$\mu$K and use the maximum of the distribution to label the collision energy in Fig.~\ref{fig:spin}.

\subsection{Molecular dynamics simulations}

\begin{figure}[t!]
	%\centering
	%\includegraphics[width=0.5\columnwidth]{Fig_chi2_vs_aSaT.pdf}
        \includegraphics[width=1.\columnwidth]{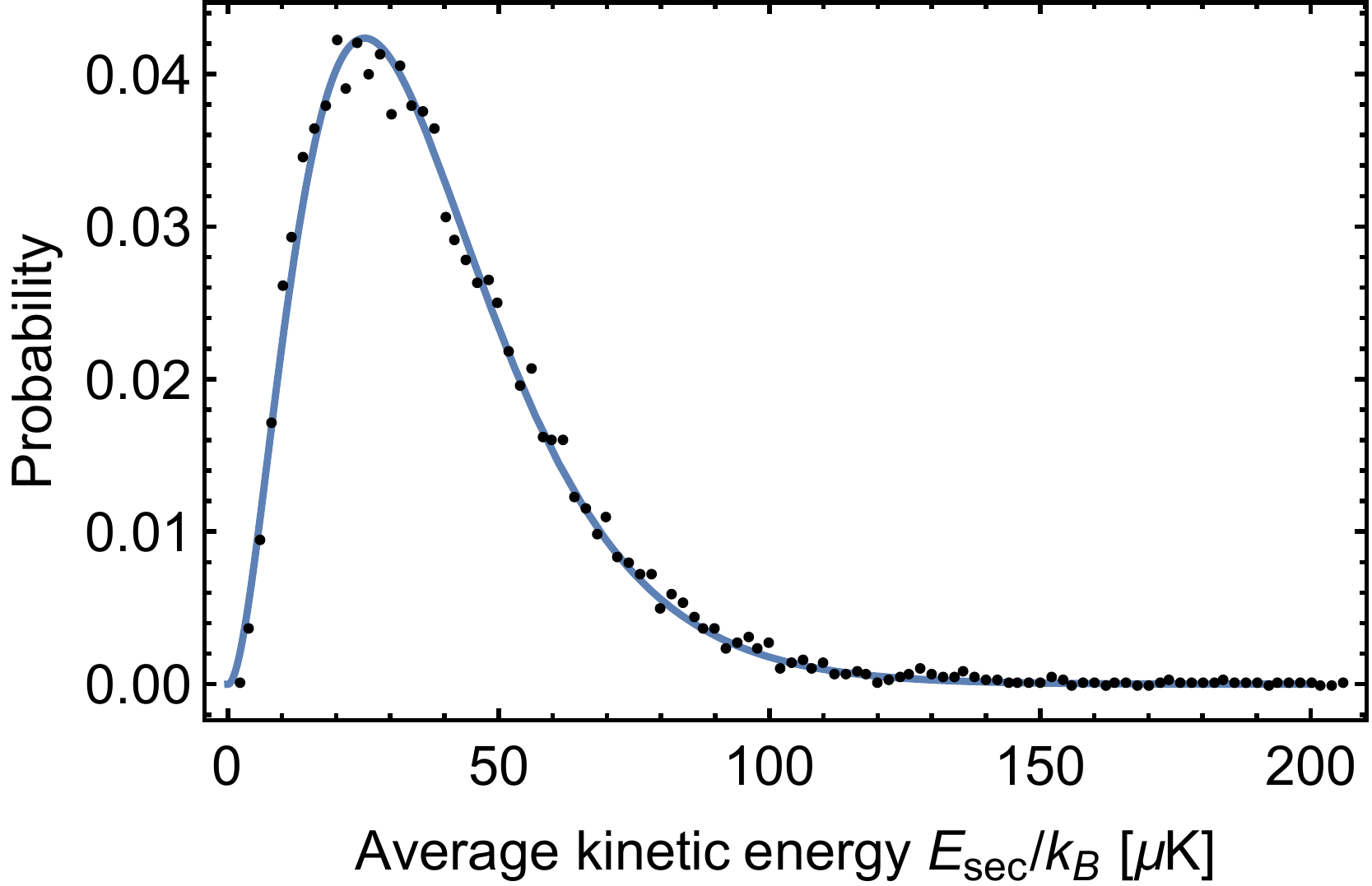}
  \caption{Calculated energy distribution after buffer gas cooling of the ion using the parameters from the experiment. The frequency of average secular kinetic energies is shown and fitted with a thermal distribution for a harmonic oscillator with a temperature of $T_{\text{sec}}^{\perp}=38.2$~$\mu$K. No significant deviation from the thermal distribution is found. The results shown are from 300 simulation runs. In these simulations, the secular kinetic energy of the ion was obtained by filtering out energy contributions with a frequency higher than half the trap drive frequency, $\Omega_{\text{rf}}/2$, as explained in Ref.~\cite{Fuerst:2018}. }
	\label{fig:histogram}
\end{figure}

We numerically simulate the full trapped-ion-atom system including the excess
micromotion detected in our experiment. To model collisions we introduce atoms
one after another at a random location on a sphere of radius $r_0=0.6\,\mu$m around
a single ion. Each atom starts with a velocity drawn from a Maxwell-Boltzmann
distribution at $T_\text{a}=10\,\mu$K and passes the sphere, where it can
interact with the ion. We set the interaction between the atom and ion
to~\cite{Fuerst:2018}:

\begin{equation}
V_\text{ia}(r)=C_4\left(-\frac{1}{2r^4}+\frac{C_6}{r^6}\right),
\end{equation}

\noindent where $C_4=5.607$~Jm$^4$ for $^6$Li/$^{171}$Yb$^+$ and we set
$C_6=5\times 10^{-19}$~m$^2$ to account for the short range repulsion
between the atom and ion. When the atom leaves the sphere, the ion's kinetic
energy (averaged over the micromotion period) is obtained, and the next atom is
introduced. We obtain the average ion cooling curve by averaging 300
simulation runs, containing $N_\text{at}=8000$ atoms each.

We fit an exponential of the form $T^{\perp}_\text{sec}(n_\text{col})
= (T_0-T_{\infty})e^{-n_\text{col}/N_\text{eq}}+T_{\infty}$ to the simulated cooling
dynamics. From this we obtain the characteristic $1/e$ number of collisions
$N_\text{eq}$ it takes to equilibrate and the final ion temperature $T_\infty$.
From the average atomic flux $\phi_\text{at} = N_\text{at}/t_\text{prop}$ through the sphere
within the total propagation time $t_\text{prop}$, we can
translate $N_\text{eq}$ into the $1/e$ number of Langevin collisions
$N_\text{L,eq}$ via
\begin{equation}
  N_\text{L,eq} = 2\pi\rho_\text{sim}\sqrt{\frac{C_4}{\mu}}\frac{N_\text{eq}}{\phi_\text{at}}\,.
\end{equation}
Here, $\rho_\text{sim} = 1/(4/3 \pi r_0^3)$ is the atomic density
in the simulation.  From comparing with the experimental $1/e$ cooling time
$\tau_\text{exp} = 244(24)\,$ms, we can deduce the atomic density $\rho_\text{at}$ in the
experiment via the Langevin rate
\begin{align}
  \Gamma_\text{L}
  = \frac{N_\text{L,eq}}{\tau_\text{exp}}=2\pi\rho_\text{at}\sqrt{\frac{C_4}{\mu}}\,,
\end{align}
to be $\rho_\text{at}=24(3)\times10^{15}\,\text{m}^{-3}$, which is in agreement with the results from absorption imaging. .

In the experiment, the buffer gas
cooling is competing with ion heating caused by electric field noise.
Independent measurements give a heating rate of $\gamma_\text{heat}=83(50)$~$\mu$K/s
in the radial direction in the absence of atoms. We account for this heating by
setting $dT^{\perp}_\text{sec}(t)/dt = -\gamma_\text{cool}
T^{\perp}_\text{sec}(t)+T_{\infty}\gamma_\text{cool}+\gamma_\text{heat}$, which
results in $T^{\perp}_\text{sec}(t) = (T_0-T_{\infty})e^{-\gamma_\text{cool}
t}+T_{\infty}+\gamma_\text{heat}/\gamma_\text{cool}$ with
$\gamma_\text{heat}/\gamma_\text{cool}\sim$~20~$\mu$K.

Finally, we obtain the energy distribution of secular ion motion from the numerical simulations. It has been found~\cite{DeVoe:2009,Zipkes:2011,Chen:2014,Rouse:2017,Weckesser:2015,Meir:2016} that the energy distribution of a trapped ion can deviate significantly from a thermal distribution when interacting with a buffer gas. In our calculations we do not find an observable difference with a thermal distribution for the secular motion of the ion after buffer gas cooling, as shown in Fig.~\ref{fig:histogram}. We attribute this result to the large mass ratio between the atoms and ion~\cite{Rouse:2017,Meir:2016}.

\subsection{Quantum scattering calculations}

\begin{figure}[t!]
	%\centering
	%\includegraphics[width=0.5\columnwidth]{Fig_chi2_vs_aSaT.pdf}
        \includegraphics[width=0.7\columnwidth]{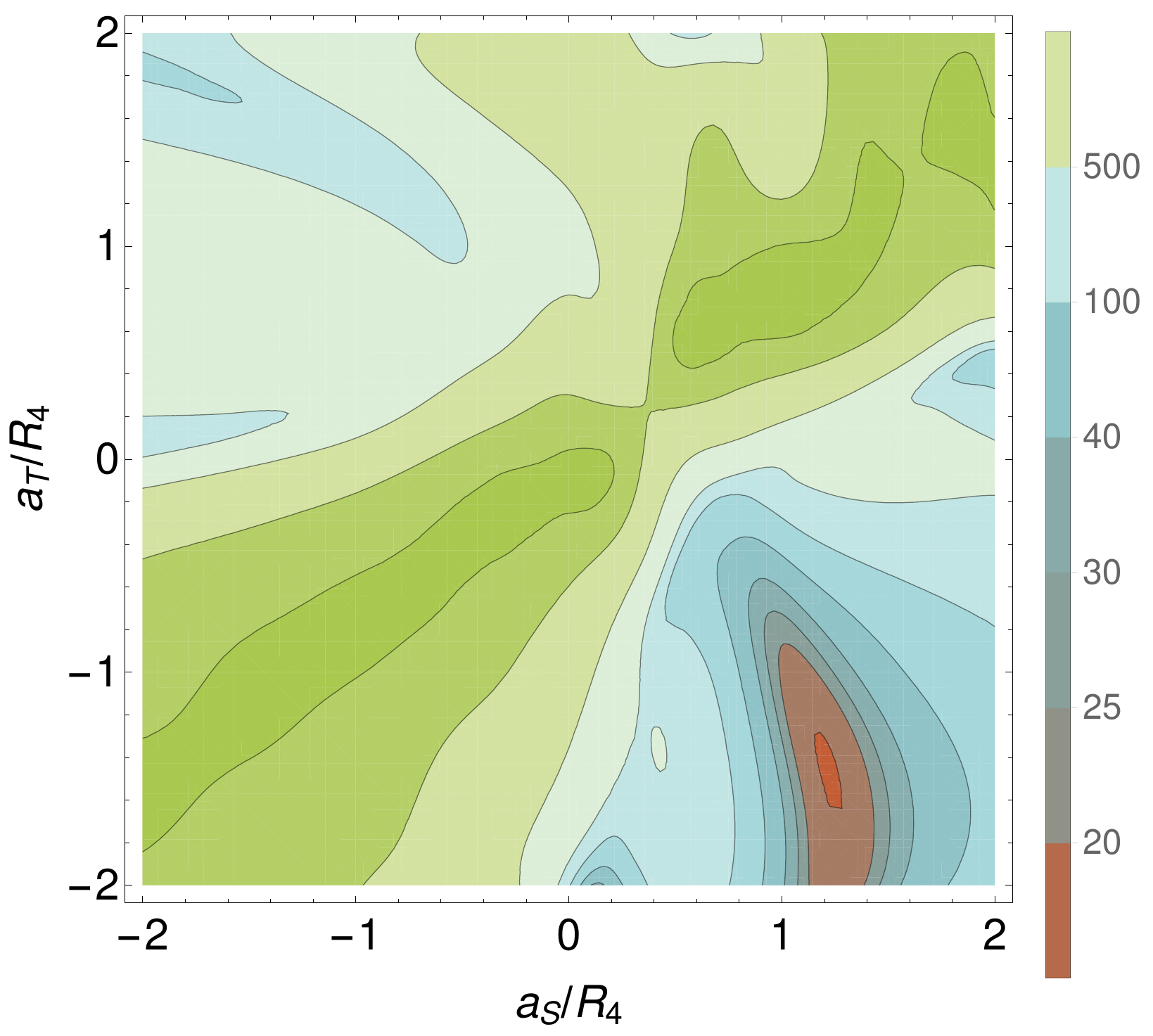}
  \caption{$\chi^2$ as a function of the singlet $a_\text{S}$ and triplet
  $a_\text{T}$ scattering lengths with the number of Langevin collisions
  optimized for each set of scattering lengths.}
	\label{fig:chi2}
\end{figure}

We construct and solve a quantum microscopic model of cold atom-ion interactions
and collisions based on the \textit{ab initio} multi-channel description of the
Yb$^+$-Li system as we presented in Refs.~\cite{Tomza:2015,Joger:2017}. The
Hamiltonian used for the nuclear motion accounts completely for all relevant degrees
of freedom including the singlet and triplet molecular electronic states, the
molecular rotation, and the hyperfine and Zeeman interactions. Experimental
values of relevant parameters including the magnetic field of 4$\,$G are
assumed. We construct the total scattering wave function in a complete basis set
containing electronic spin, nuclear spin and rotational angular momenta.

We solve the coupled-channels equations using a renormalized Numerov
propagator %~\cite{Johnson:1978}
with step-size doubling. The wave function ratios
are propagated to large interatomic separations, transformed to the diagonal
basis, and the $K$ and $S$ matrices are extracted by imposing the long-range
scattering boundary conditions in terms of Bessel functions. As an entrance
channel, we assume Yb$^+$ in the $|F=1,m_F=-1\rangle$ state and Li in the
$|F=1/2,m_F=-1/2\rangle$ or $|F=1/2,m_F=1/2\rangle$, while all allowed other
channels are included in the model. The inelastic rate constants and scattering
lengths are obtained from the elements of the $S$ matrix summed over relevant
channels including different partial waves~$l$.

We calculate the rate constant for spin-changing collisions
$K(E,a_\text{S},a_\text{T})$ as a function of the singlet $a_\text{S}$ and
triplet $a_\text{T}$ scattering lengths. The scattering lengths of the singlet
and triplet potentials are fixed by applying uniform scaling factors $\lambda_i$
to the interaction potentials: $V_i(R)\to\lambda_i V_i(R)$. We express
scattering lengths in units of the characteristic length scale for the atom-ion
interaction $R_4=\sqrt{\mu C_4/\hbar}$. Next, the rate constant is convoluted
with the ion's energy distribution induced by a controlled micromotion added to
a thermal offset of $E_0/k_{\text{B}}$=20$\,\mu$K
\begin{equation}
\bar{K}(\bar{E},a_\text{S},a_\text{T})=\int_{E_0}^{E_0+\bar{E}_\text{eMM}}P_{\bar{E}_\text{eMM}}(E-E_0) K(E,a_\text{S},a_\text{T}) dE\,.
\end{equation}
The probability of detecting the ion spin in $|F=0,m_F=0\rangle$ after preparing
it in $|F=1,m_F=-1\rangle$ is calculated
\begin{equation}
S(\bar{E},a_\text{S},a_\text{T},n_\text{L}) = 1 - \exp(- n_\text{L} \cdot  \bar{K}(\bar{E},a_\text{S},a_\text{T}) / K_\text{L})\,.
\end{equation}
where $n_\text{L}$ is the number of Langevin collisions and
$K_\text{L}=2\pi\sqrt{C_4/\mu}$ is the Langevin collision rate coefficient. The
singlet and triplet scattering lengths are found together with the number of
Langevin collisions by minimizing the $\chi^2$ function
\begin{equation}
\chi^2(a_\text{S},a_\text{T},n_\text{L})=\sum_{i=1}^{N_\text{exp}} \left(\frac{S_\text{exp}(\bar{E}_i)-S(\bar{E}_i,a_\text{S},a_\text{T},n_\text{L})}{\sigma_i}\right)^2\,,
\end{equation}
which quantifies how well our theoretical model reproduces the measured values
$S_\text{exp}(\bar{E}_i)$.
The numerical minimization of $\chi^2$ yields $a_\text{S} = 1.2(0.3)R_4$,
$a_\text{T} = -1.5(0.7)R_4$, and $n_\text{L} = 1.2(0.4)$, where the
uncertainties of the theoretical values are obtained by imposing that $\chi^2$
give the $p$-value equal or better than 0.05. Figure~\ref{fig:chi2} shows
corresponding $\chi^2$ dependence on the scattering lengths.

\section*{Acknowledgements}
This work was supported by the European Union via the European Research Council
(Starting Grant 337638) and the Netherlands Organization for Scientific Research
(Vidi Grant 680-47-538, Start-up grant 740.018.008 and Vrije  Programma 680.92.18.05) (R.G.). D.W.~and M.T.~were supported by the National Science
Centre Poland (Opus Grant~2016/23/B/ST4/03231) and PL-Grid Infrastructure. We thank Jook Walraven and Corentin Coulais for comments on the manuscript.

%\section*{Author contributions}
%T.F. and R.G. conceived the experiment. T.F., H.F., H.H., N.V.E. and M.M. performed the experiment. H.F. and R.G. performed molecular dynamics simulations, D.W. and M.T. performed quantum scattering simulations. All authors contributed to discussions about
%the experiment, the analysis of the data and the preparation of the manuscript.

%\section*{Data availability}
%The data that support the plots within this paper and other findings of this study are available from the  corresponding author upon reasonable request.

%\section*{Code availability}
%The code that supports the plots within this paper and other findings of this study is available from michal.tomza@fuw.edu.pl and the the corresponding author upon reasonable request.

%\section*{Competing interests}
%The authors declare no competing interests

\section*{References}

%-------------------------------
%\bibliography{biblio-RG}

\end{document}